\documentclass[manuscript]{aastex}
\usepackage{graphicx}

\newcommand{\kms}[1]{$#1$~km~s$^{-1}$}

\shorttitle{MMF}
\shortauthors{Lim et al.}

\begin{document}

\title{First Simultaneous Detection of Moving Magnetic Features in Photospheric Intensity and Magnetic Field Data}

\author{Eun-Kyung Lim, Vasyl Yurchyshyn, and Philip Goode}
\affil{Big Bear Solar Observatory, New Jersey
Institute of Technology, 40386 North Shore Lane, Big Bear City, CA
92314-9672, USA}
\email{eklim@bbso.njit.edu}

\begin{abstract}
The formation and the temporal evolution of a bipolar moving magnetic feature (MMF) was studied with high spatial and temporal resolution. The photometric properties were observed with the New Solar Telescope at Big Bear Solar Observatory using a broadband TiO filter ($705.7$~nm), while the magnetic field was analyzed using the spectropolarimetric data obtained by Hinode. For the first time, we observed a bipolar MMF simultaneously in intensity images and magnetic field data, and studied the details of its structure. The vector magnetic field and the Doppler velocity of the MMF were also studied. A bipolar MMF having its positive polarity closer to the negative penumbra formed being accompanied by a bright, filamentary structure in the TiO data connecting the MMF and a dark penumbral filament. A fast downflow ($\leq$ \kms{2}) was detected at the positive polarity. The vector magnetic field obtained from the full Stokes inversion revealed that a bipolar MMF has a U-shaped magnetic field configuration. Our observations provide a clear intensity counterpart of the observed MMF in the photosphere, and strong evidence of the connection between the MMF and the penumbral filament as a serpentine field.
\end{abstract}


\section{INTRODUCTION}
Moving magnetic features (MMFs) were first detected in CN spectroheliograms by \citet{She69} as small bright {features} moving radially outward from sunspots at speeds of around \kms{1}. Although these observations were not accompanied by magnetic field measurements, the authors suggested that these bright moving features could be the manifestation of magnetic field erupting through the surface. Subsequently, \citet{Har73} found that those moving bright features represent bipolar magnetic concentrations, which they named ``moving magnetic features". Based on the observation that MMFs only appear around decaying sunspots, \citet{Har73} interpreted MMFs as a cross section of twisted magnetic flux tube that is detached from the sunspot magnetic field and poking through the surface. Although it was later shown that MMFs are not necessarily always associated with decaying sunspots \citep{Wan91, Mar02}, \citet{Har73}'s interpretation has been widely accepted in follow-up studies \citep[etc.]{Vra74, Yur01, Zha03, Hag05}.

More recent observations, however, showed that MMFs are extensions of the sunspot penumbral field \citep[etc.]{Hag05, Sai05, Rav06, Cab06, Cab07, Cab08, Sai08}. \citet{Lee92} reported that the majority of MMFs are formed close to the dark penumbral filaments that are thought to consist of more horizontal magnetic field components. \citet{Yur01} showed that the orientation of MMFs is not random but strongly correlated with both the sign and the amount of the large-scale twist of the sunspot magnetic field. This result indicates that magnetic field of MMFs is part of the sunspot magnetic field, and at least some of observed MMFs could be explained with the serpentine field model of \citet{Har73}. The possible connection between MMFs and the penumbral field was also addressed in a theoretical study by \citet{Sch02}, who suggested that MMFs may be a part of a serpentine field, which is a prolonged horizontal penumbral field carrying the Evershed flow. Observations by \citet{Sai05} supported this idea by showing the subsequent formation of MMFs along the same magnetic filaments extending from the penumbra into the moat surrounding a sunspot. \citet{Cab06, Cab07, Cab08} provided a strong observational evidence that some MMFs are magnetically connected to penumbral field by showing that observed MMF formed as a continuation of the penumbral Evershed flow. The Evershed flow is known to mainly occur along the horizontal magnetic component of the penumbral field \citep{Tit93, Sta97}. The horizontal component of the penumbral field was observed to return back into the convection zone, accompanying a fast downflow at the opposite polarity region where the field returns back into the sun \citep{Wes97, Wes01, Sch00, Bel04, Ich07}, and such observations were supported by the numerical simulations \citep{Kit10, Rem11}. Based on observed downflows associated with MMFs near the penumbral edge, it has been suggested that the Evershed flow may play an important role in MMF formation by pushing inclined field into surface making a U-loop configuration \citep{Zha07, Shi08}.

There were efforts to study the intensity signatures of MMFs in the photosphere, which are not yet fully explored partially owing to the spatially unresolved small-scale nature of MMFs. In most studies, MMFs were often observed as tiny moving bright features \citep{She69}, or ``moving H$\alpha$ emission clouds" \citep{Har73}. Using white-light sunspot data, \citet{Shi87} observed bright, granule-like features traveling outward from the penumbral boundary at \kms{2.7}. Although their association to MMFs was not observationally established, their travel speed is in agreement with that of certain MMFs moving as fast as \kms{2-3} \citep{Shi01}. Simultaneous observation of MMFs in G-band and Ca~{\footnotesize II}~K spectral lines showed that MMFs were observed as bright features in both the photosphere and chromosphere \citep{Ryu97}. \citet{Bon04, Bon05} carried out further studies on bright features in G-band and showed that most of them form adjacent to dark penumbral filaments and then move outward from the sunspot at a speed of \kms{0.7}. The statistical study by \citet{Hag05} showed that MMFs follow preferred paths that were consistent with the moat flows around sunspots. Recently, \citet{Lim11} studied granule-like features that formed at the edge of dark penumbral filaments associated with MMFs, and detected a slight brightening in both TiO and H$\alpha$ filtergrams. Some theoretical models of MMFs explained the appearance of bright features in terms of the temperature difference along the serpentine field \citep{Ryu97,Sch02}.

In spite of the great progress in understanding of MMFs, essential details of their fine structure and their exact connection to the penumbral field still remain elusive mainly because of their small spatial scale ($<1\arcsec$). For example, recent data acquired by \citet{Lim11} with the high-resolution New Solar Telescope (NST) at Big Bear Solar Observatory (BBSO) showed that the photospheric features, associated with MMF evolution, reveal even more complex sub-structures. High-resolution photometric data from the NST complemented by high-spatial, high-sensitivity spectropolarimetric data will contribute to the understanding the relationship between MMFs and the penumbral field, and thus the origin of MMFs.

\section{OBSERVATIONS AND DATA ANALYSIS}
\begin{figure}[tb]
\begin{center}
    \includegraphics[width=0.6\textwidth]{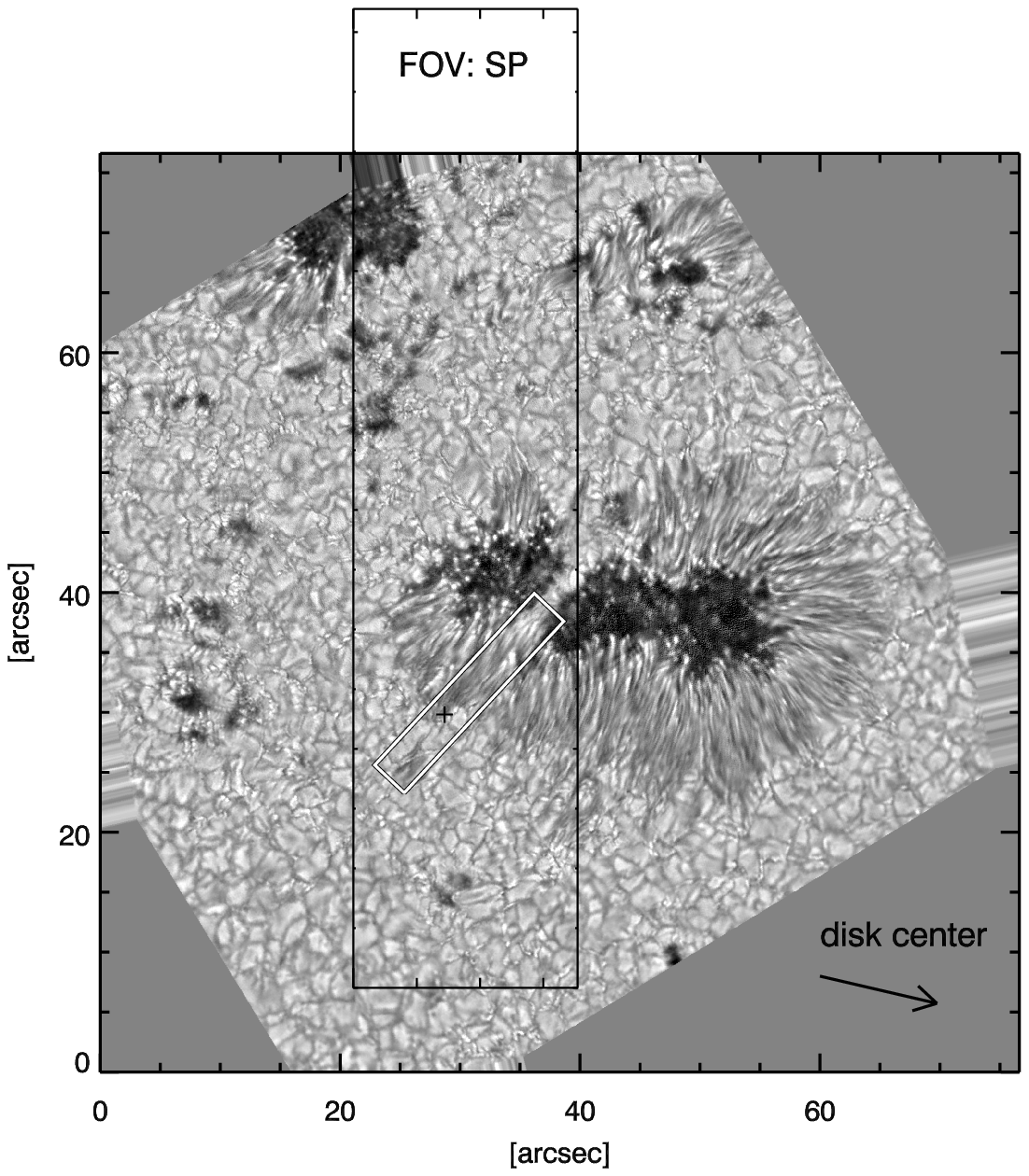}
    \includegraphics[width=1\textwidth]{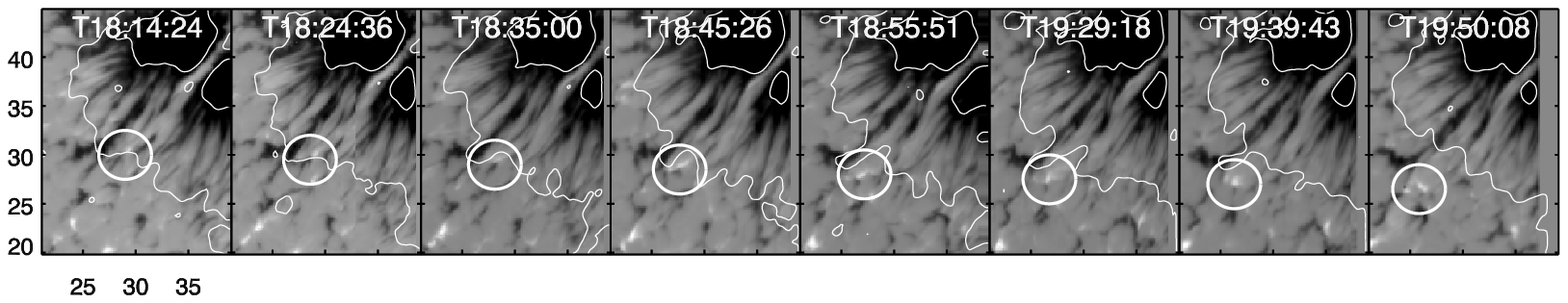}
    \caption{Active region NOAA 11283 being located at N$12$E$25$. \emph{Top}: the FOV of SOT/SP is overplotted on the NST/TiO image. The white rectangle indicates the FOV of the slice where the MMF of interest was studied in detail. The arrow indicates the direction to disk center. \emph{Bottom}: SOT/SP Line-of-sight (LOS) magnetograms obtained from ME inversion. Magnetograms are saturated at $\pm 1$~kG. Our interested positive magnetic patch associated with an adjacent negative polarity is indicated by circles. Its position at 18:14:24~UT in the FOV of TiO data is presented by $+$ symbol in the top panel. White contours represent moat/penumbra and penumbra/umbra boundaries defined as ${I}/{\overline{I}}$ to be $1$ and $0.68$, respectively, where $\overline{I}$ is mean intensity of SOT/SP.}\label{fov}
\end{center}
\end{figure}

Active region (AR) NOAA 11283 first appeared on the east solar limb on 2011 August 30, and was simultaneously observed by both BBSO/NST and the Solar Optical Telescope \citep[SOT;][]{Sue08,Tsu08} on board Hinode \citep{Kos07}, from 18:24~UT to 19:39~UT on 2011 September 03. The leading negative polarity of the AR consisted of a well-developed sunspot, while the following positive polarity fields were comprised of a number of pores. The main sunspot, in particular the outer boundary of the penumbral filament (indicated by `$+$' symbol in Figure~\ref{fov}), is the object of our study. It was located at N$12$E$25$ when the NST/TiO observations began, which corresponds to the $\mu$-value of $0.906$. The NST broadband filter imager \citep{Cao10b} with a TiO filter ($705.7$~nm) took a burst of 100 images every 15~s with an exposure time of $1.2$~ms. The TiO data covered $76\arcsec \times 76\arcsec$ field of view (FOV), with a pixel size of $0\arcsec.0375$, while the diffraction limit of the NST is $0\arcsec.11$ at the wavelength of the TiO line. The NST operates with the aid of an adaptive optics system \citep{Cao10a} under normal seeing conditions. All TiO data were first speckle reconstructed using the Kiepenheuer-Institut Speckle Interferometry Package \citep{Wog08}, then de-stretched to remove residual image distortion.

Hinode/SOT was operated using both the narrowband filter imager (NFI) and spectro-polarimeter \citep[SP;][]{Ich08}, which targeted AR~11283 from 18:13:30~UT to 20:00:00~UT. The SOT/NFI took Stokes I and V maps in the Na~D spectral line ($589.6$~nm) with the FOV of $123\arcsec \times 131\arcsec$ and the spatial sampling of $0\arcsec.16$ every 64~s. Full Stokes I, Q, U, and V profiles were obtained at two photospheric Fe~I lines ($630.15$ and $630.25$~nm) from SOT/SP scanning. For the SOT/SP observations, the normal mapping mode was used, which produces a spatial sampling of $0\arcsec.16$. The scanning FOV was reduced to $19\arcsec \times 82\arcsec$ in order to achieve a higher time cadence of $10.4$~min. We thus obtained $8$ SP scans, and $6$ of them were taken simultaneously with the TiO data (except the first and the last scan). Both SOT/NFI and SOT/SP data were calibrated using standard SolarSoft routine. In order to obtain the vector magnetic field, we applied the Milne-Eddington (ME) inversion \citep{Kat11} to all Stokes profiles, then resolved the $180^{\circ}$ ambiguity using the method introduced by \citet{Moon03}. Since a penumbra is believed to consist of magnetic field that is inhomogeneous both horizontally and vertically \citep{Sol93, Tit93a, Mar00, Bel04}, physical values derived from the ME inversions must be considered as an average of various components along the atmosphere with non-negligible uncertainties. The line-of-sight (LOS) velocity, $v_{\textrm{\tiny LOS}}$, was measured independently by utilizing a center-of-gravity (COG) method \citep{Sem67,Sem70,Ree79}. It was shown that the COG method derives the LOS velocity accurately even in the case of asymmetric lines. In case of strong downflows or upflows, on the other hand, the velocity may be underestimated by the COG method \citep{Uit03}. Since we focus on the outer edge of penumbral filaments, where strong downflows are often observed, it is possible that the measured value of the downflow speed was underestimated.

All data sets from different instruments were carefully co-aligned with each other, and presented in the top panel in Figure~\ref{fov}. SOT/NFI I maps were first co-aligned with NST/TiO images, then the SOT/SP vector magnetograms were co-aligned with SOT/NFI V/I maps. As seen in the figure, the sunspot had an irregular shape with its well-developed penumbral structure not yet fully wrapped around the sunspot, and a well-defined light bridge splitting the umbra into two parts. The one-hour TiO data set showed many interesting phenomena of filamentary structure formation at the outer tip of penumbral filaments. One of those events occurred well-inside the region where FOVs of both NST/TiO and \emph{Hinode}/SOT instruments overlapped (indicated by `$+$' symbol in Figure~\ref{fov}). We thus selected a narrow region along the direction of the penumbral filament, where a bipolar MMF was well-observed from its origin (inside circles in the bottom panels). This slice, outlined by the white rectangle in the figure, is at an angle of $47^{\circ}$ to the solar east-west direction, and \textbf{$60^{\circ}$} angle to the direction toward the disk center. After co-alignment, all vector magnetograms were corrected for the projection effect to obtain the magnetic component vertical to the solar surface, ${B}_z$, horizontal to the surface, ${B}_{xy}$, the inclination angle, $\gamma$, and the azimuth angle with respect to the slice orientation. In order to derive the vertical velocity, $v_z$, to the solar surface, the LOS velocity, $v_{\textrm{\tiny LOS}}$, was corrected for the projection effect assuming the transverse velocity to be zero. Since the real value of the transverse velocity component may not be zero, this assumption may result in the underestimation of the $v_z$.

\section{RESULTS}

\begin{figure}[tb]
\begin{center}
    \includegraphics[width=1\textwidth]{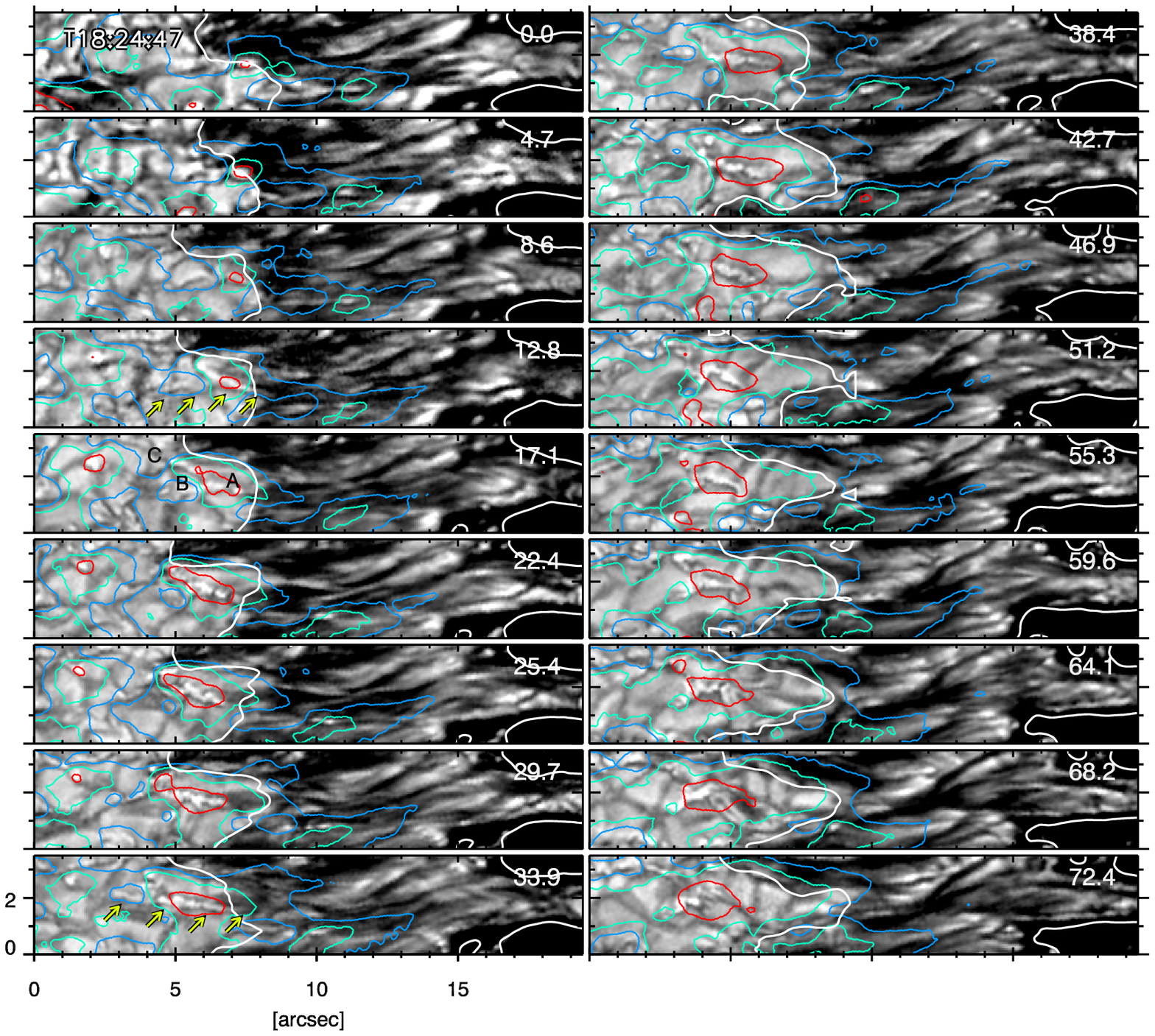}
    \caption{TiO series of images showing evolution of the MMF. The observation time of the first image is presented in UT on the upper-left corner, and then the elapsed time is shown in minutes. Blue, green and red contours indicate the SOT/NFI LOS magnetic strength of $-150$, $-50$, and $100$~G, respectively. White contours represent moat/penumbra (left-hand side) and penumbra/umbra (right-hand side) boundaries defined as ${I_{\textmd{\scriptsize TiO}}}/{\overline{I}_{\textmd{\scriptsize TiO}}}$ to be $1$ and $0.68$, respectively, where $\overline{I}_{\textmd{\scriptsize TiO}}$ is mean intensity of TiO data (same in Figure~\ref{allslice}). Yellow arrows in the $t=12.8$ and $t=33.9$ frames indicate the bright, filamentary feature of our interest.}\label{tio-slice}
\end{center}
\end{figure}

Figure~\ref{tio-slice} shows the observed photospheric filamentary structure that appeared and protruded from the outer edge of a penumbral filament. We selected each TiO image from the slice region, and overplotted SOT/NFI V/I maps multiplied by the calibration coefficient $\alpha=12344$ to convert the polarization signal to magnetic flux density. The calibration coefficient $\alpha$ was determined by comparing SOT/NFI V/I with SOT/SP vector magnetograms obtained from the ME inversion. Contours with blue, green and red colors correspond to $-150$~G, $-50$~G and $100$~G, respectively. The tick values along the slit direction (x-axis) increase toward the umbra. The object of our interest is a tiny but well-pronounced bright, elongated structure, which is co-spatial with the red contour at (7.5, 2) in the 18:24:47~UT frame. The feature was in the immediate vicinity of the outer boundary of the penumbra (represented by the white contour in each panel). Although it was not yet prominent at this time, its round shape is clearly distinct due to the darker surroundings. The bright feature became elongated and its filamentary shape was well observed from $t=12.8$ to $t=42.7$. One of the most important characteristics is that it formed emerged from, and was connected to the dark penumbral filament (indicated by yellow arrows in the $t=12.8$ and $t=33.9$ frames). The feature maintained its filamentary structure connected to the dark penumbral filament for about 50 min, and was detached from the filament at $t=51.2$, when a granular cell developed between them. The width of the filamentary structure was about $0\arcsec.5$ and the length was about $4\arcsec$ at $t=38.4$.

Another interesting characteristic of the filamentary structure is that it was co-spatial with the positive magnetic patch throughout the time it grew and traveled outward from the penumbra. Figure~\ref{tio-slice} shows that this positive magnetic patch also grew in size and traveled outward from the penumbra. Its size was $0\arcsec.5$ at first then it developed into a $3\arcsec$ feature in the last panel. Figure~\ref{tio-slice} also shows that this positive magnetic patch (marked with \textbf{A}, in $t=12.8$ panel) evolved synchronously with a negative magnetic patches (\textbf{B} and \textbf{C}) on the farther side from the penumbra \citep{Yur01,Sai08}. Careful analysis of the TiO image reveals that the bright filamentary structure split into two branches and was connecting not only \textbf{A} and \textbf{C}, but also \textbf{A} and \textbf{B} ($t=17.1$). The bottom panels in Figure~\ref{fov} clearly show that \textbf{A} and \textbf{B} were moving as a pair (inside circles). The elongated shape of the positive magnetic patch became more roundish after the co-spatial filamentary structure detached from the penumbra ($t=72.4$). The measured properties of the positive magnetic patch, such as estimated averaged velocity (\kms{0.66}), the size ($3\arcsec$), are consistent with \citet{Sai08}, who studied bipolar MMFs. They also observed that the elongated shape of the magnetic patches became more roundish outside the spot. Therefore, the observed positive magnetic patch could be considered as an inner foopoint of a type I MMF \citep{Shi01}, and hence the TiO filamentary structure represents an intensity counterpart of the observed MMF. And the alignment of the negative penumbra, positive magnetic patch and the associated negative polarity along the bright, filamentary structure suggests the possible connection with each other. These unique data show, for the first time, even finer structure of the intensity counterpart of MMFs in the photosphere at high-spatial and high-temporal resolution. The clear connectivity between the observed bipolar MMF and the dark penumbral filament seen in TiO intensity data also supports previous findings showing that some MMFs form as the continuation of the penumbral field \citep[etc.]{Har73, Sch02, Hag05, Sai05, Sai08, Cab06, Cab07, Cab08}.

\begin{figure}[tb]
\begin{center}
    \includegraphics[width=0.41\textwidth]{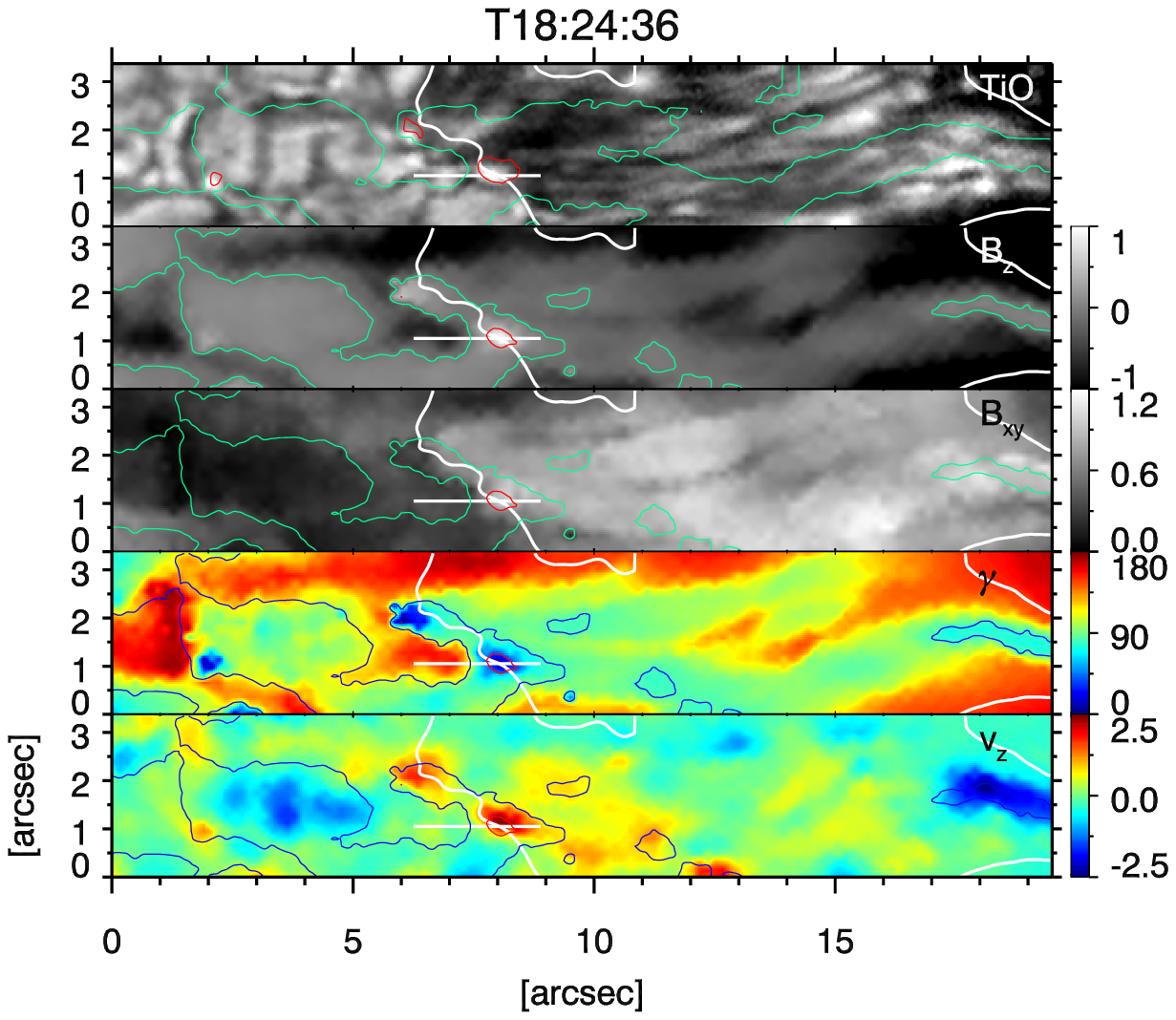}
    \includegraphics[width=0.41\textwidth]{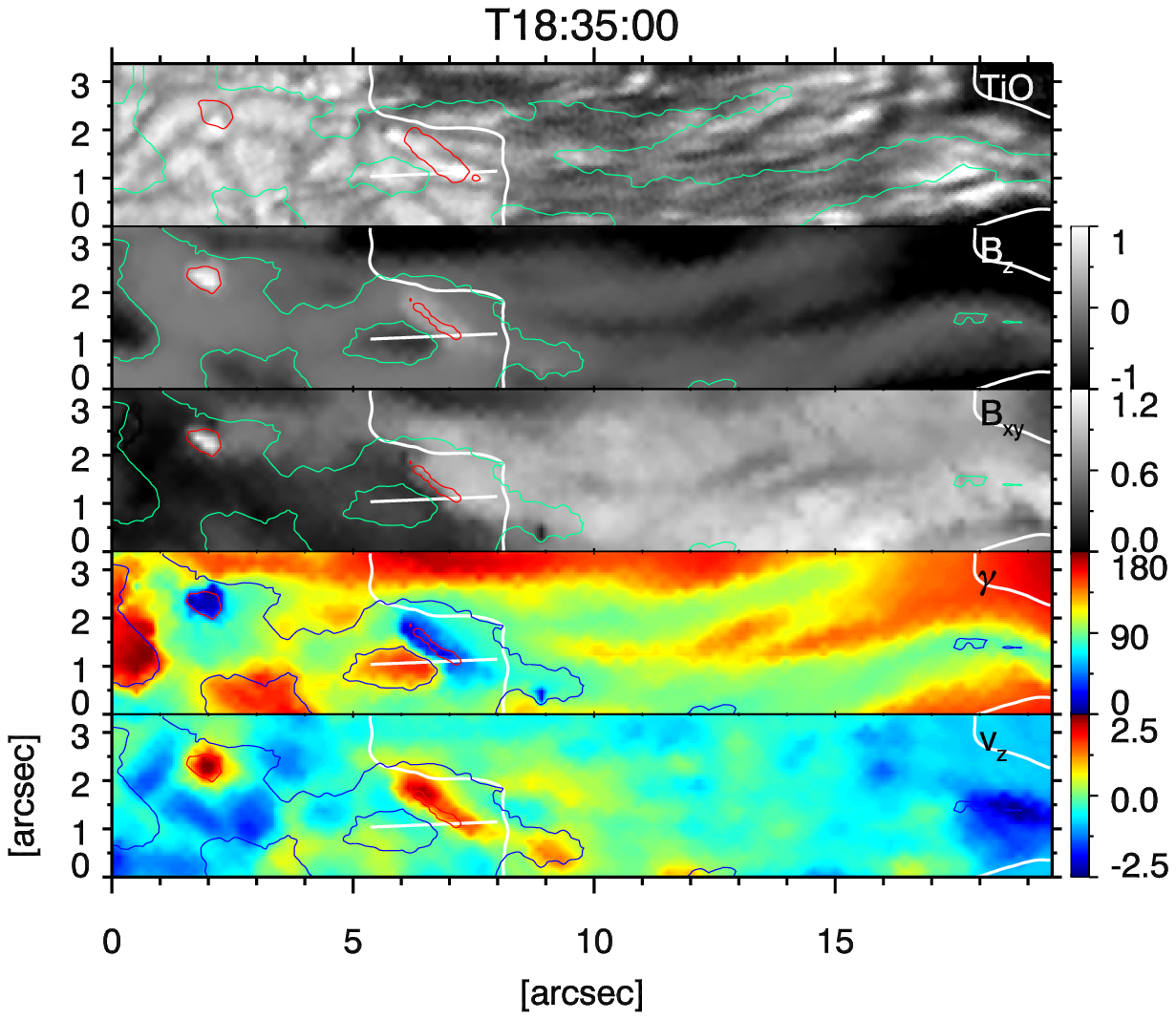}\\
    \includegraphics[width=0.41\textwidth]{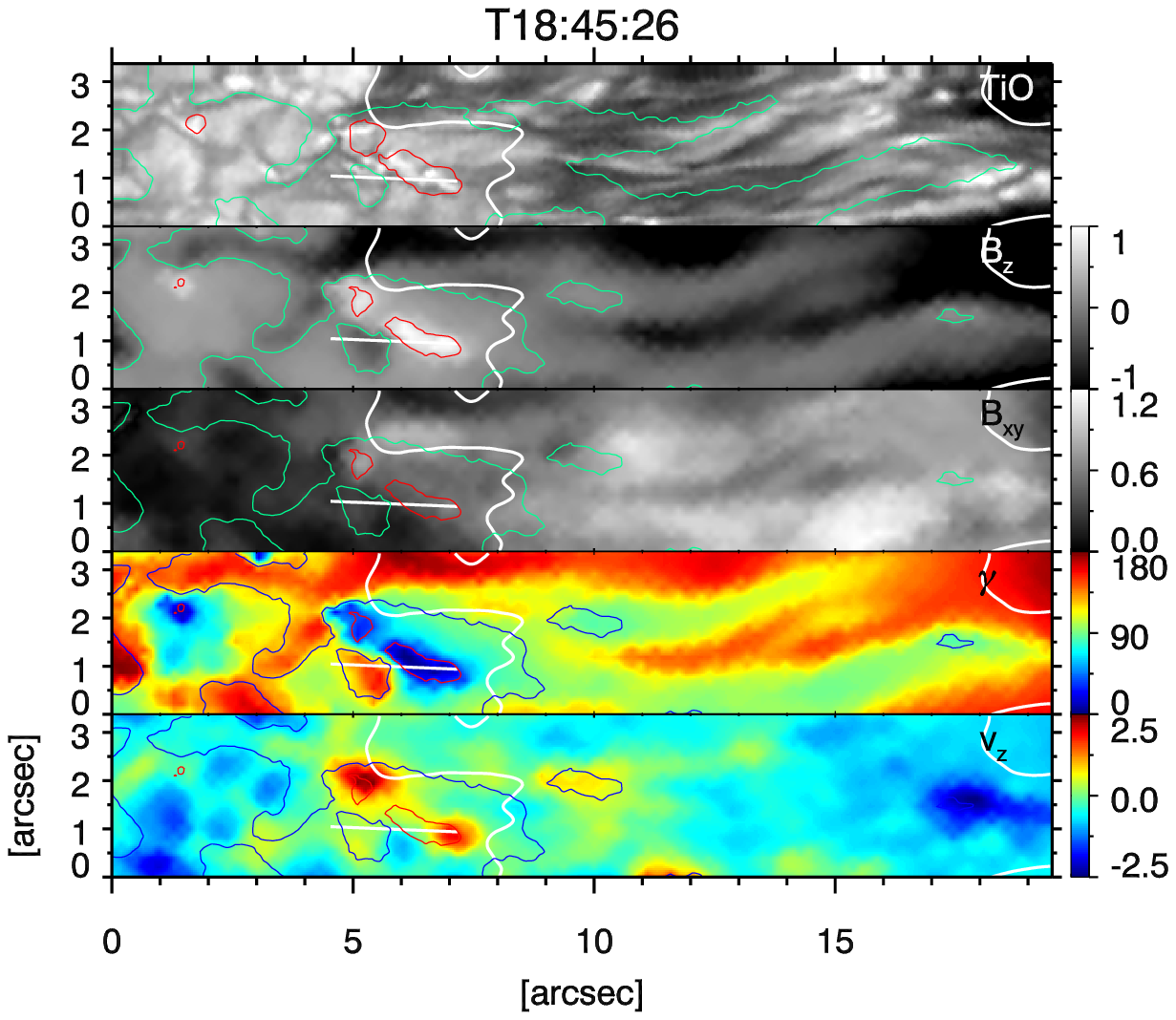}
    \includegraphics[width=0.41\textwidth]{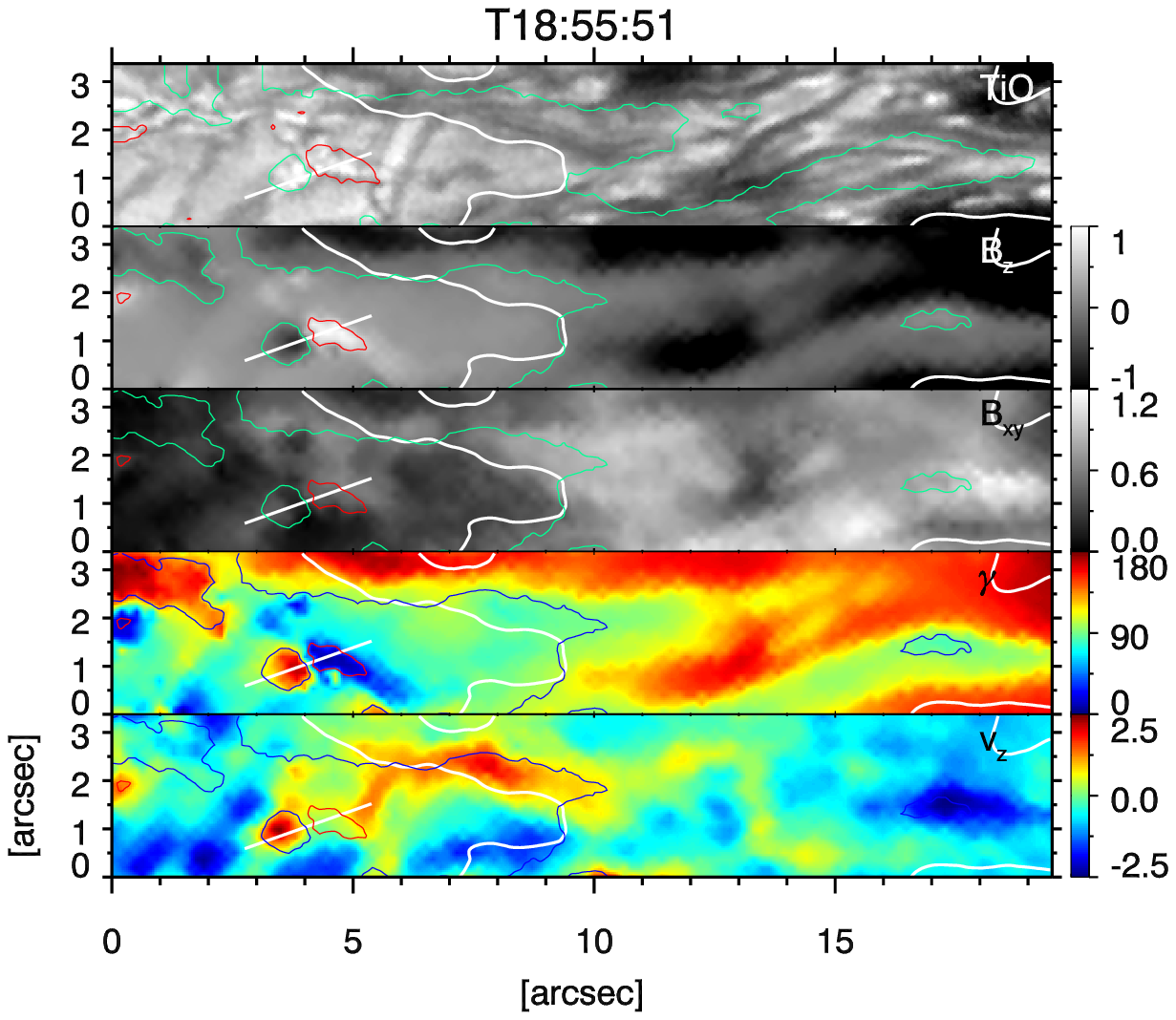}\\
    \includegraphics[width=0.41\textwidth]{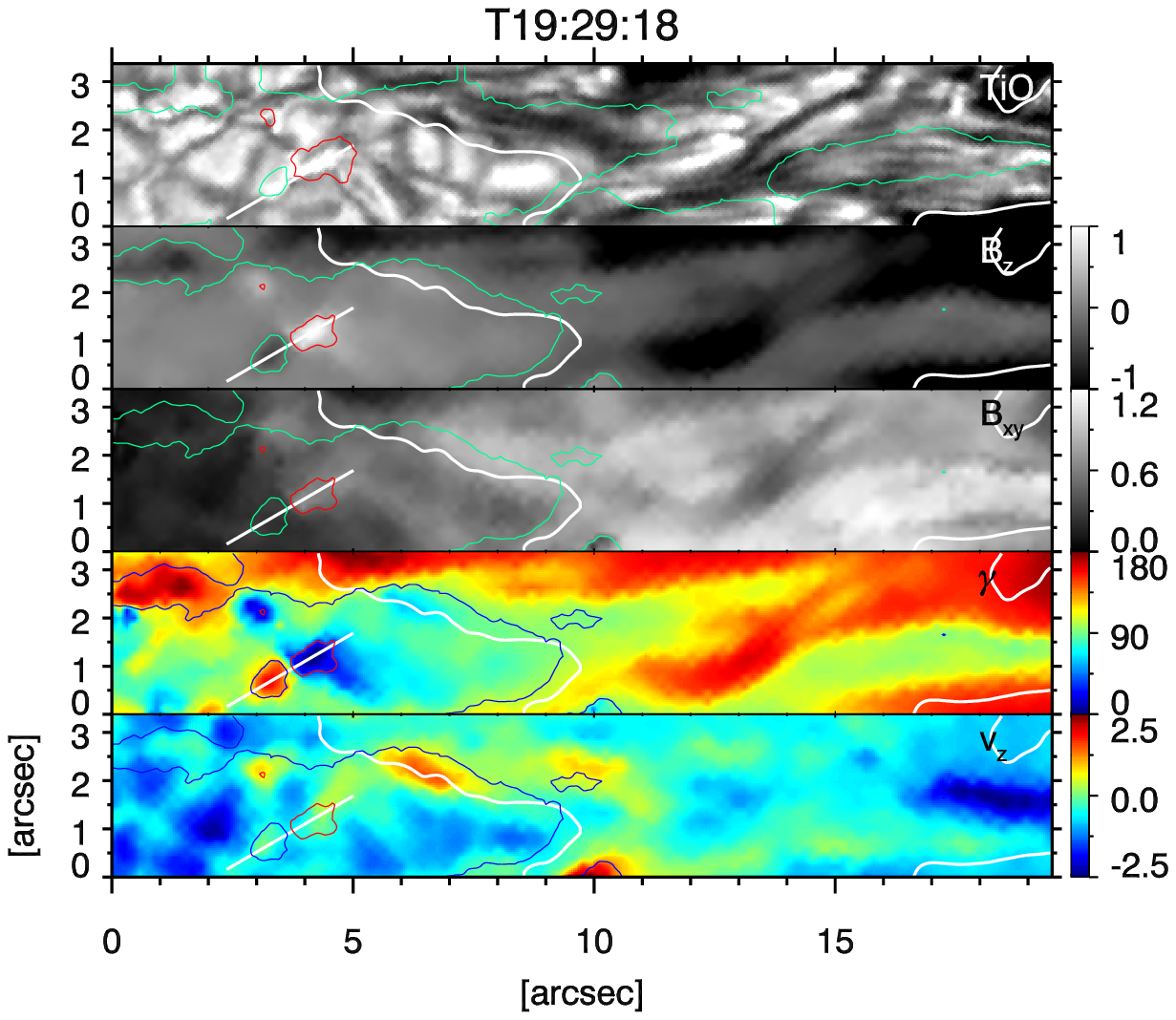}
    \includegraphics[width=0.41\textwidth]{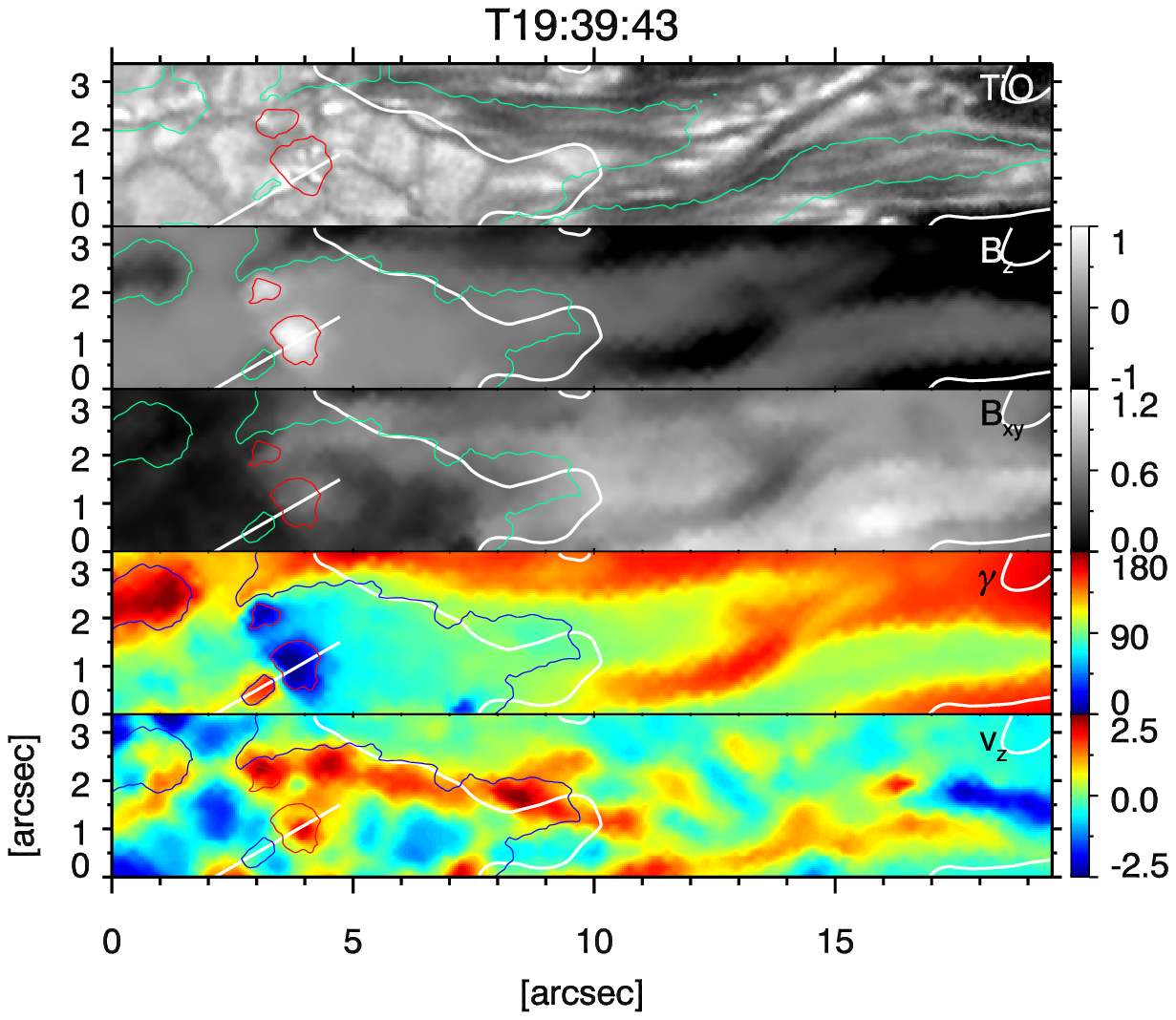}\\
    \caption{Each panel shows NST/TiO, vertical component of vector magnetic field to the local frame obtained from ME inversion, ${B}_{z}$, (scaled between $\pm 1$~kG), transverse component of vector magnetic field to the local frame, ${B}_{xy}$, ($0$ to $1.2$~kG), inclination angle ($0^{\circ}$ to $180^{\circ}$), and the vertical velocity field $v_{z}$ ($\pm$ \kms{2.5}). Overplotted contours represent ${B}_{z}$ (Line-of-sight magnetic strength over TiO image only) values at $-200$~G (green or blue), and $+300$~G (red) levels.}\label{allslice}
\end{center}
\end{figure}

In order to understand the magnetic field structure of the observed MMF, we analyzed vector magnetic field data around the MMF region. Figure~\ref{allslice} shows time series of the SOT/SP data carefully co-aligned with NST/TiO images. Only 6 out of 8 SOT/SP magnetograms were taken simultaneously with NST/TiO data and displayed in Figure~\ref{allslice}. Both magnetic and velocity data were corrected for the projection effect except magnetic contours overlaid on the TiO images. The vertical magnetic field, ${B}_{z}$, clearly shows that the MMF of interest has a bipolar configuration having the opposite magnetic polarity to that of the sunspot at the inner footpoint. The horizontal magnetic field strength, ${B}_{xy}$, decreases between the positive and the negative magnetic patch (18:24:36~UT -- 18:35:00~UT), implying that the positive magnetic patch is a part of the horizontal penumbral field at this time and the field becomes more vertical beyond the inner footpoint of the MMF. The inclination angle of $0^{\circ}$ and $180^{\circ}$ corresponds to the positive and the negative magnetic field, respectively. The ``uncombed" \citep{Sol93} nature of the penumbra is well-illustrated in the inclination maps. The comparison between the inclination maps and the TiO images shows that the dark penumbral filament consists of highly inclined (horizontal) magnetic field (panels 18:24:36~UT -- 18:45:25~UT). Recall that the positive magnetic patch was in the region where the horizontal magnetic field strength decreased, and the bright, filamentary structure was observed mostly co-spatial with the positive magnetic patch. Interestingly, the velocity maps show that the relatively strong downflow component ($\sim$ \kms{2}) is detected at the positive magnetic patch, especially at the early phase of the MMF formation (18:24:36~UT -- 18:45:25~UT). Similar downflow events near the outer boundary of a penumbra have been reported in the past \citep{Wes01,Shi08}, and they were interpreted as the outer bound of the Evershed flow due to the turning of the penumbral field into the convection zone.

\begin{figure}[tb]
\begin{center}
    \includegraphics[width=0.24\textwidth]{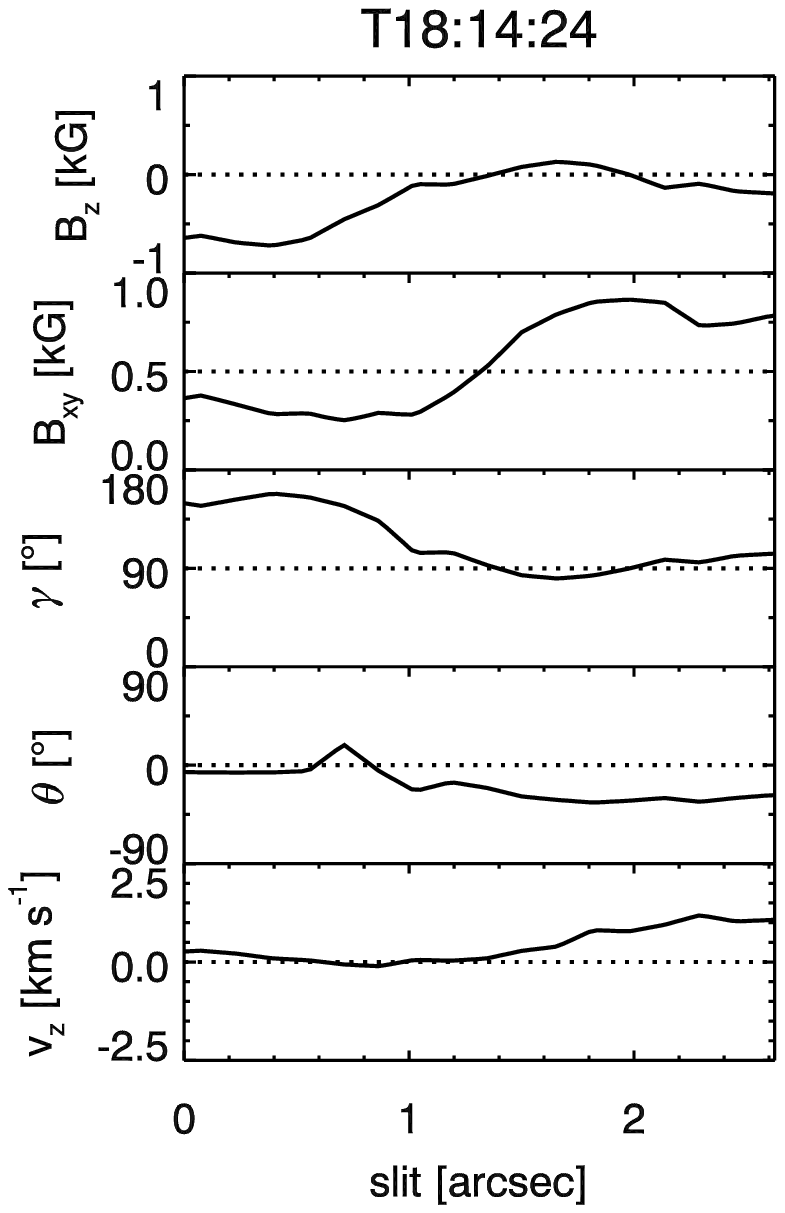}
    \includegraphics[width=0.24\textwidth]{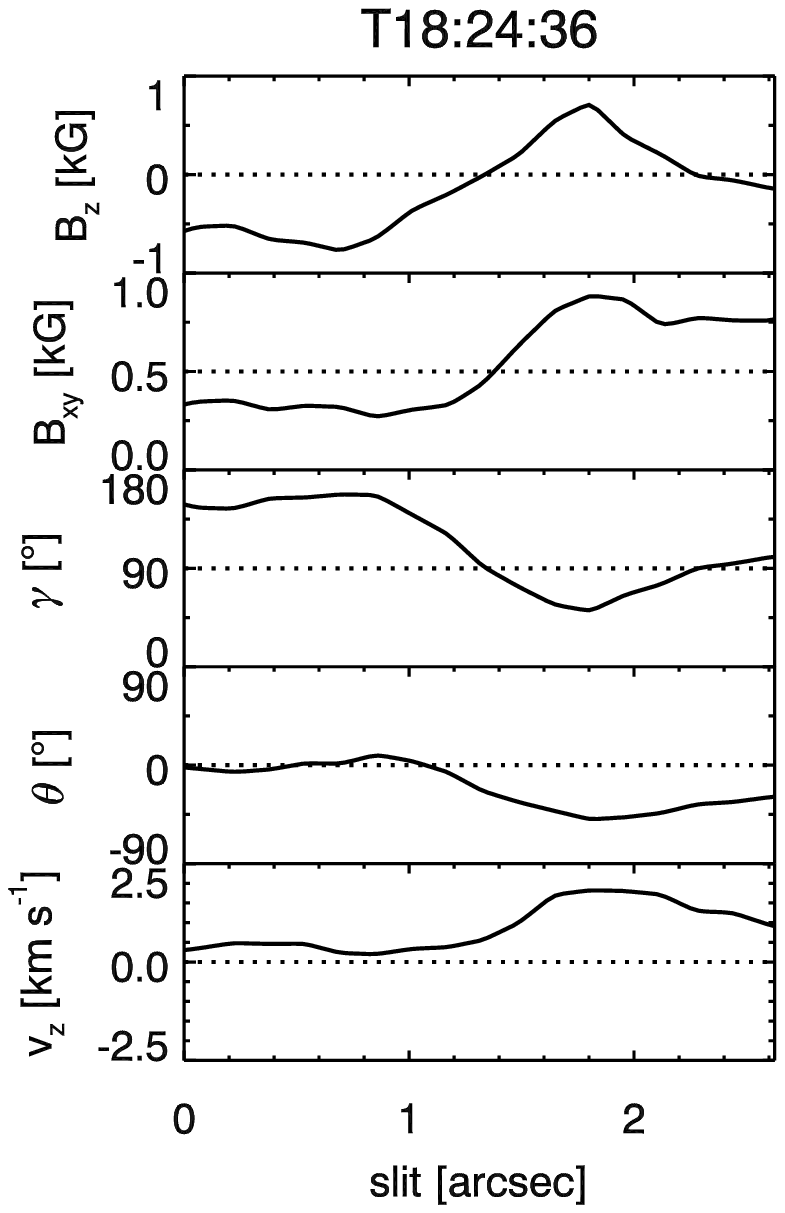}
    \includegraphics[width=0.24\textwidth]{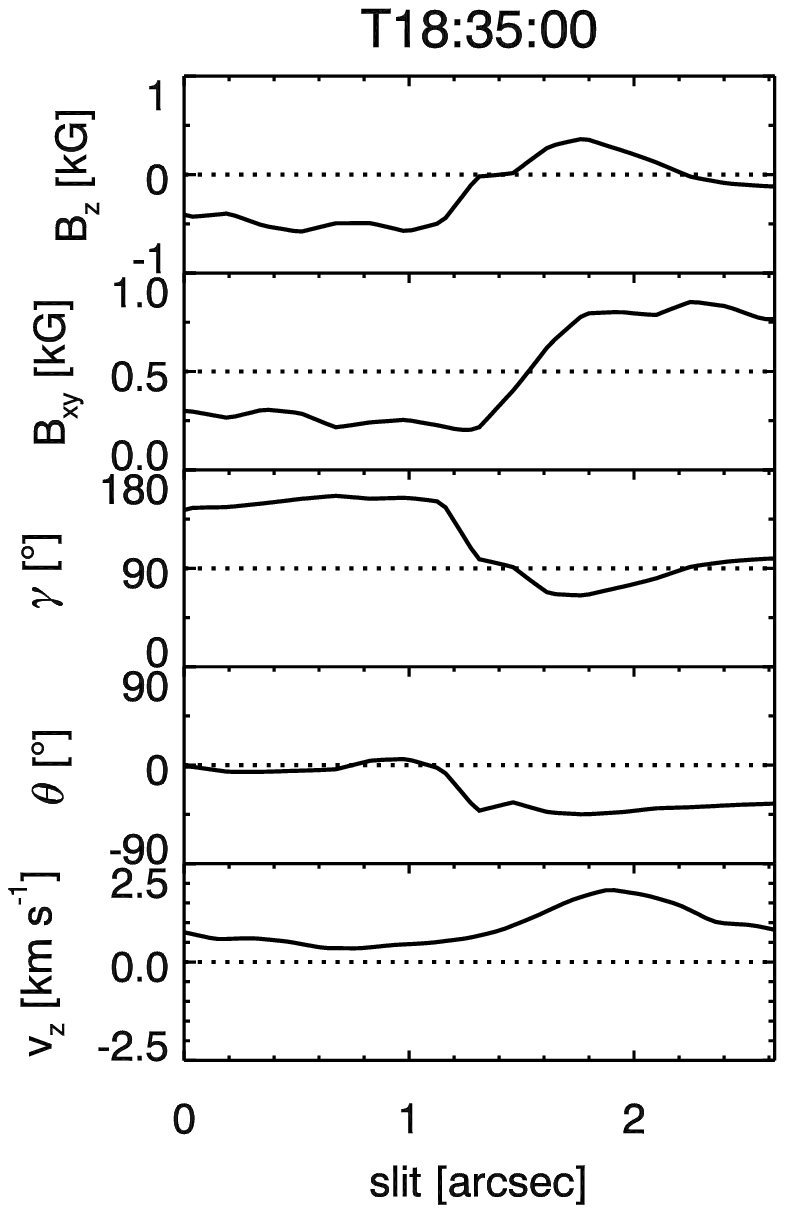}
    \includegraphics[width=0.24\textwidth]{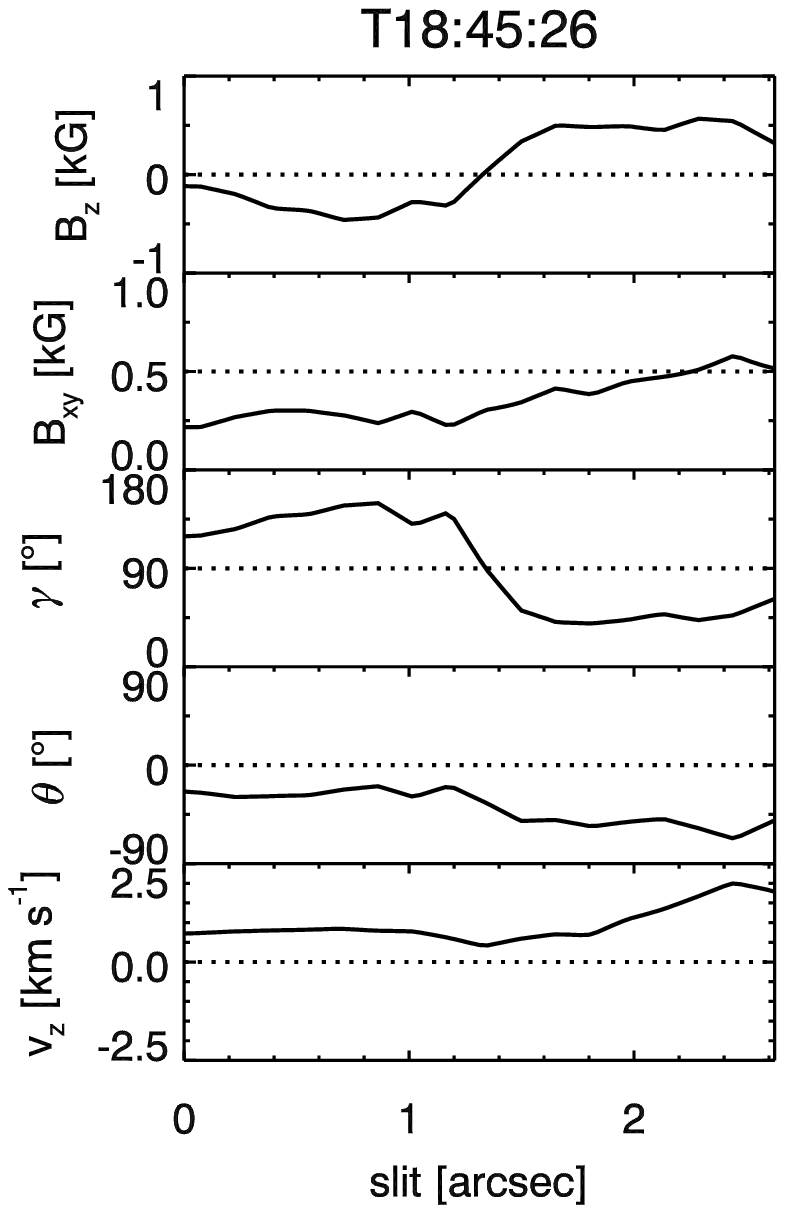}\\
    \includegraphics[width=0.24\textwidth]{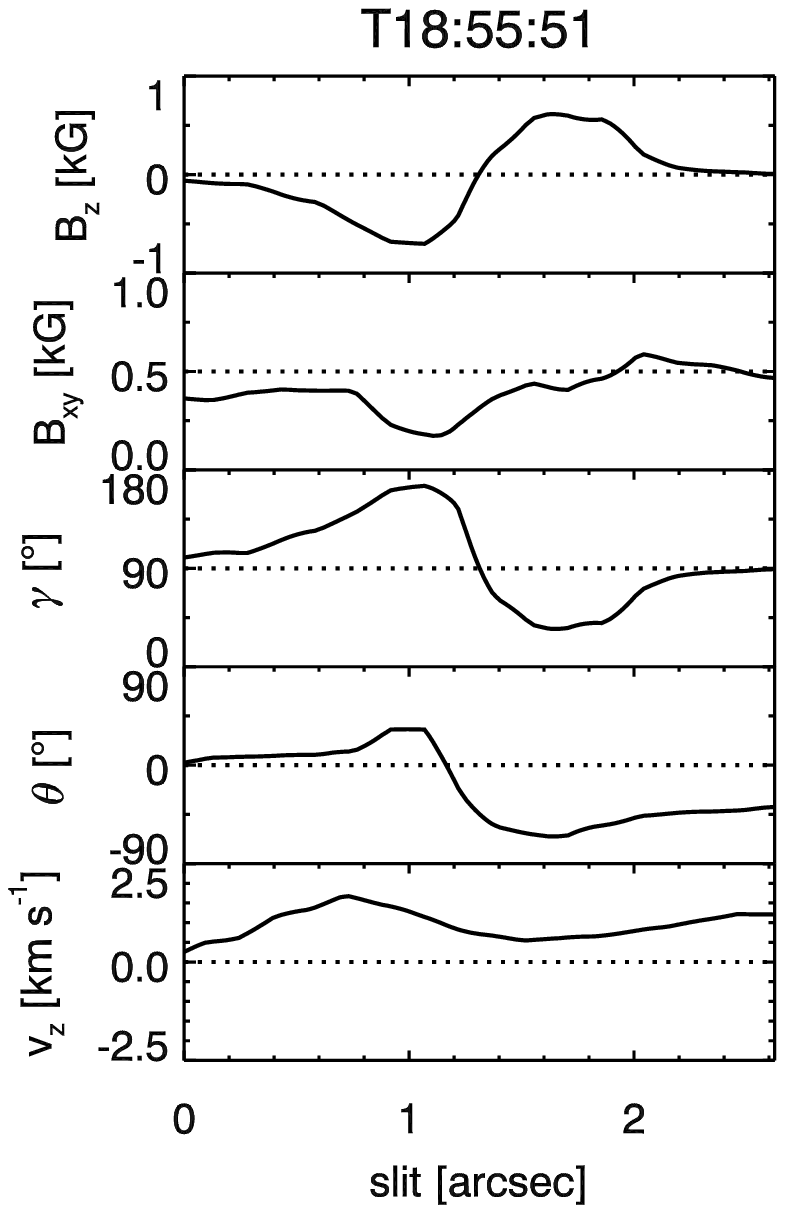}
    \includegraphics[width=0.24\textwidth]{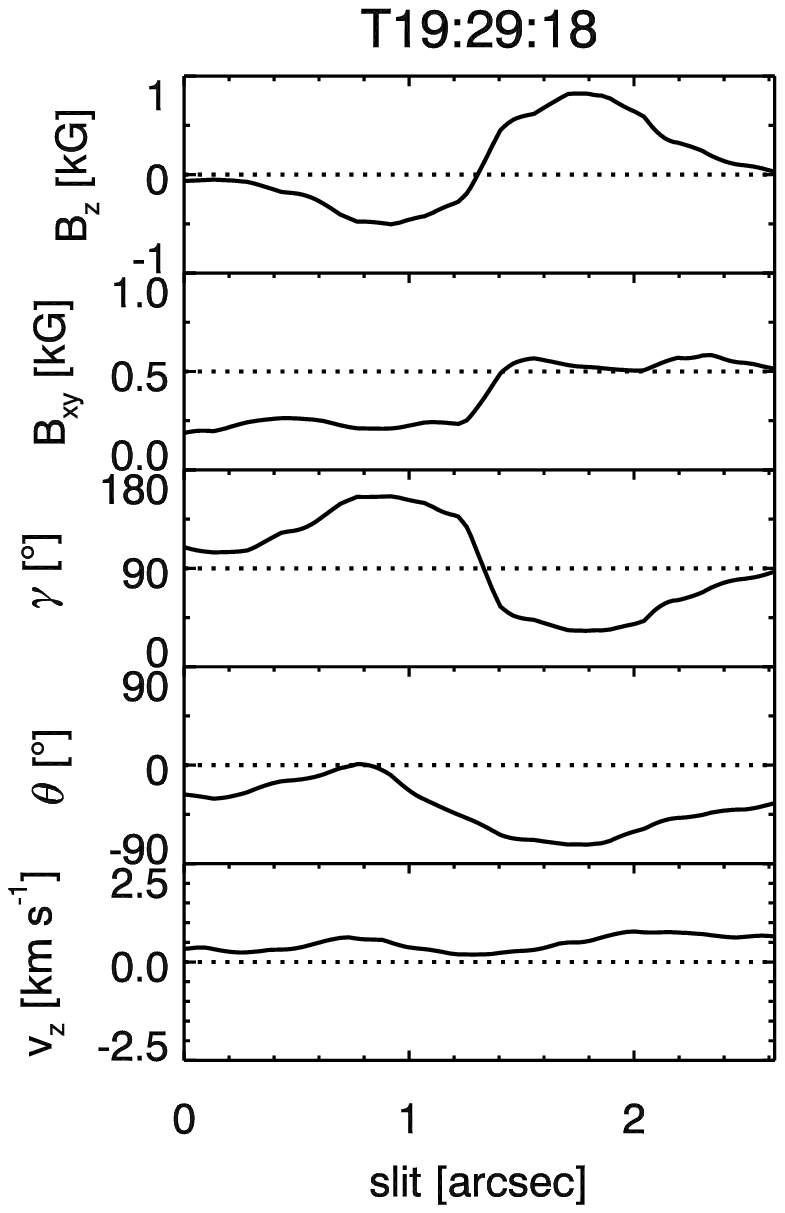}
    \includegraphics[width=0.24\textwidth]{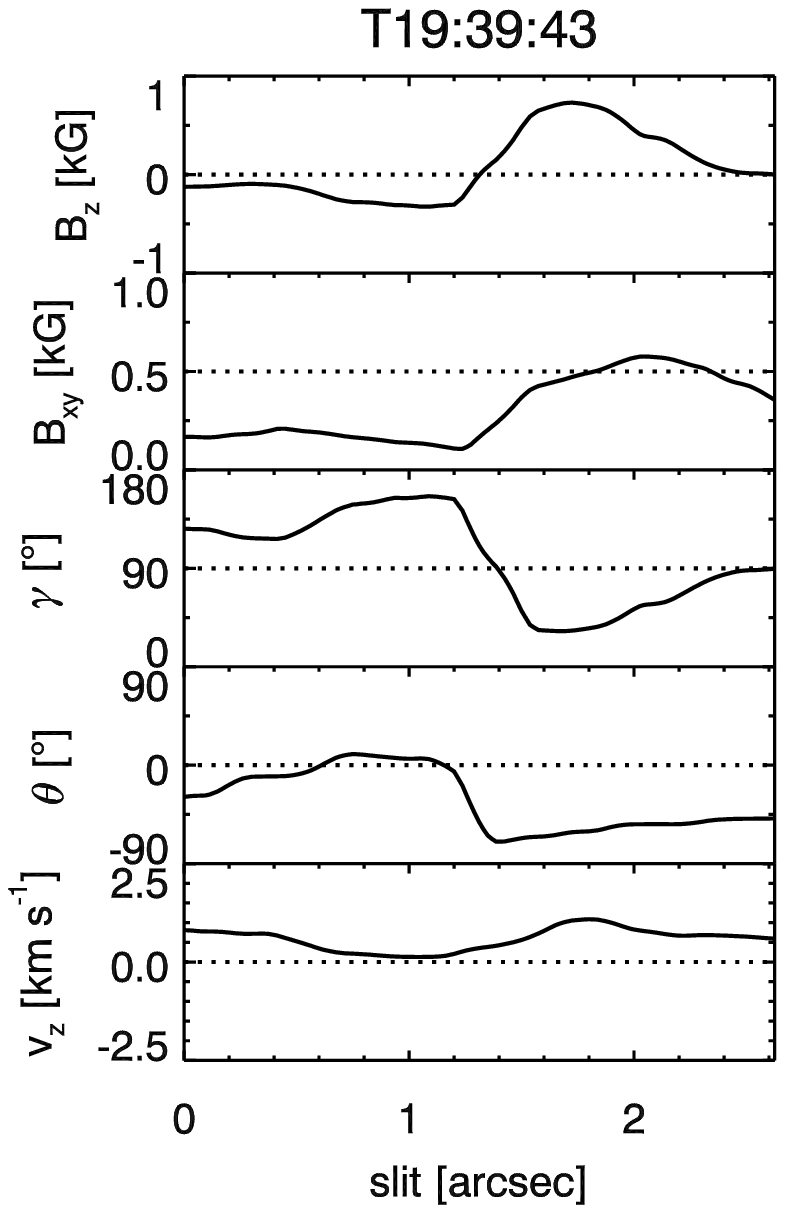}
    \includegraphics[width=0.24\textwidth]{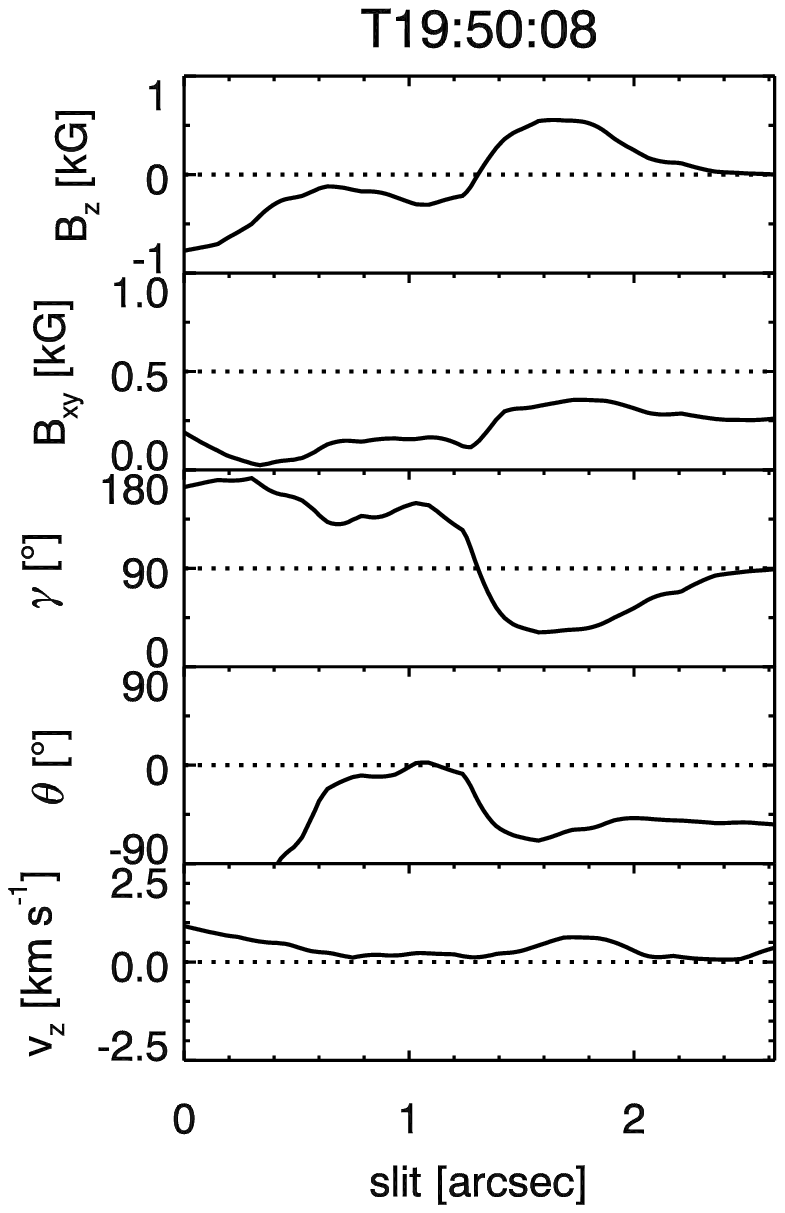}
    \caption{Profiles of ${B}_{z}$, ${B}_{xy}$, inclination angle, $\gamma$, azimuth angle, $\theta$, and the projection corrected vertical velocity component, ${v}_z$, across the opposite magnetic polarities of the MMF. All physical quantities are measured along the slit shown in Figure~\ref{allslice}.}\label{profiles:new}
\end{center}
\end{figure}

\begin{figure}[tb]
\begin{center}
    \includegraphics[width=1\textwidth]{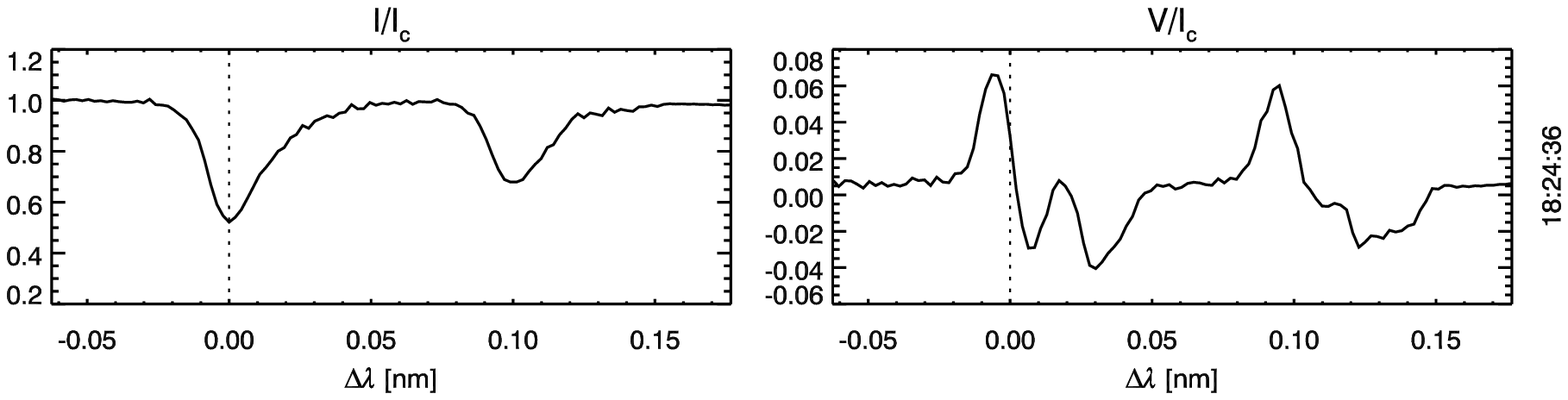}
    \caption{Stokes I and V profiles normalized by continuum intensity, I$_c$, were obtained at the position where $v_z$ showed its greatest value on the positive magnetic patch at 18:24:36~UT.}\label{profiles}
\end{center}
\end{figure}

In order to explore spatial and temporal variations of the magnetic and velocity fields in the MMF, we produced spatial profiles of the SOT/SP data following the slit running along the axis of the bipolar MMF (white line segments in Figure~\ref{allslice}; note that all 8 SOT/SP datasets were used in this case). The profiles were generated for the vertical component of magnetic field, ${B}_z$, horizontal magnetic field strength, $B_{xy}$, inclination angle, $\gamma$, azimuth angle with respect to the slit, $\theta$, and the vertical component of the velocity, $v_z$ (Figure~\ref{profiles:new}). At 18:14:24~UT, the positive magnetic patch begins to form at $x=1.8$ and the field inclination nearly $90^{\circ}$ indicates that the magnetic field was nearly horizontal at this time. The profiles for $B_{xy}$ and $\gamma$ show that the weak positive magnetic patch appeared in the region, where the $B_{xy}$ decrease starts and the field gradually turns from horizontal to vertical in the direction of the decreasing abscissa, which is farther away from the umbra. The $v_z$ profile shows a weak downflow ($\sim$ \kms{1}) in the region of the weak positive patch at this time. Later, at 18:24:36~UT, field became more vertical in the region of the positive magnetic patch, and the downflow speed increased up to $\sim$ \kms{2}. The profiles of ${B}_z$, $B_{xy}$, $\gamma$ along with the $v_z$, show that the strengthening of the positive magnetic patch and the vertical magnetic field profile occurred along with the continuous existence of a downflow. The peak vertical magnetic field strength reached up to $-700$~G (at 18:55:51~UT) and $800$~G (at 19:29:18~UT) in the negative and the positive polarity, respectively and their separation was estimated to be $\sim 0.8\arcsec$. The field azimuth, $\theta$, was oriented mostly along the slit direction pointing toward the umbra. We thus conclude that the observed bipolar MMF appears to have the U-loop magnetic field configuration, and the positive magnetic patch, inner footpoint of the MMF, is the region where the penumbral magnetic field returns back into the surface. The strong downflow may be interpreted as a vertical component of the Evershed flow along the return penumbral field \citep{Sch00, Wes01, Bel04, Rem11}. We also found abnormal lobes in the Stokes V profile inside the positive magnetic patch (Figure~\ref{profiles}), where strong downflow was detected, similar to \citet{Cab06}. The abnormal lobes indicate that the magnetic and the velocity fields have complex stratifications along the LOS direction, and hence both the magnetic and the velocity values derived from the ME inversion in this region should be carefully interpreted as averaged values along the atmosphere.

\section{CONCLUSIONS AND DISCUSSIONS}
We simultaneously observed the formation of a bipolar MMF with NST/TiO filter imager, Hinode/NFI and Hinode/SP. The high spatial and temporal resolution of the NST provided us unique intensity data on the evolving MMF at the photospheric level. From a 2-hour data set, we obtained the following primary results.

\begin{enumerate}
\item The newly forming bipolar MMF was accompanied by a bright, filamentary structure in the photosphere protruding from the dark penumbral filament. The filamentary structure was observed to be brighter than its surroundings in the region of the positive magnetic patch of the MMF that was closer to the negative penumbra. The connection between the filamentary structure and the dark penumbral filament was interrupted as a granular cell developed in between.
\item The vector magnetogram showed that the observed bipolar MMF has a U-loop magnetic field configuration. The inner footpoint of the MMF (positive magnetic patch) is the region where penumbral magnetic field returns back into the surface, and the farther footpoint (negative magnetic patch) consists of the field re-emerging through the surface. The positive magnetic patch was associated with a strong downflow ($\leq$ \kms{2}) throughout the observational period.
\end{enumerate}

The significance of this study is that we showed for the first time the intensity counterpart of the MMF in the photosphere with high spatial resolution. It formed immediately from at the outer edge of a dark penumbral filament and evolved into a bright, filamentary structure before it appeared to be detached from the penumbral filament. After the bright structure was detached from the penumbra, it moved into and blended with an inter granular lane. The observed structure could be related to the `Evershed clouds' described in \citet{Shi94} and \citet{Cab06,Cab07}, but it is not certain if they represent the same structure. Although the size and the average proper motion \citep[$2250$~km, \kms{0.5};][]{Cab06} are consistent with our observation ($0.5$\arcsec\, by $4$\arcsec, \kms{0.66}), Evershed clouds were reported to be associated with the Doppler blueshift and appeared in the mid-penumbra \citep{Shi94,Cab06,Cab07}, while the filamentary structure we observed was associated with the Doppler redshift. They identified Evershed clouds from the Doppler velocity enhancement, while we identified the filamentary structure from the intensity map when it appeared to be brighter than the surroundings. Besides, due to our limited time period of observations, we could not observe if the structure first appeared well-inside the penumbral boundary.

Similarly to the Evershed cloud in \citet{Cab06}, the filamentary structure studied here was co-spatial with a bipolar MMF. The inner footpoint of the bipolar MMF showed the opposite magnetic polarity to that of the sunspot, same as in \citet{Yur01,Zha03, Cab06, Sai08}, and such an opposite-polarity pair was interpreted as footpoints of U-loop magnetic field that is a part of sea-serpent field line in \citet{Sai08}. Our analysis of the vector magnetic field confirmed the U-loop magnetic field configuration of the observed bipolar MMF \citep{Zha07}. The result shows that the penumbral field returns into the surface through one footpoint of the MMF having an opposite magnetic polarity to the sunspot with the downflow component, which is in agreement with the previous studies \citep{Sch00, Wes01, Bel04, Shi08, Kit10, Rem11}. Moreover, from the inclination angle profile, we also found that the initially horizontal field component of the MMF changed to more vertical as the MMF evolved in association with the strong downflow. We thus conclude that the MMF formed as a result of the U-loop magnetic field sinking into the surface due to the downward Evershed flow. \citet{Kit10} explained that the bipolar MMFs form by the magnetic field lines `stretched by the downward flows and dragged under the surface'. This observation is also consistent with \citet{Sch02} showing that the penumbral magnetic field with strong Evershed flow is developed into a sea-serpent shape, and the same idea on the MMF formation was also supported by \citet{Zha07} and \citet{Shi08}. Moreover, the clear connection between the intensity counterpart of the observed bipolar MMF and the dark penumbral filament also strongly supports previous results showing that some MMFs form as continuations of penumbral field in the form of sea-serpent field lines \citep{Har73, Sch02, Hag05, Sai05, Sai08, Cab06, Cab07, Cab08}.

\acknowledgments Authors appreciate the anonymous referee for constructive criticisms and comments that greatly improved the manuscript. Authors are grateful to BBSO team for their contribution to the instrumentations and the observations. E.-K.L. and P.G. are partially supported by AFOSR (FA9550-12-1-0066), and V.Y. is partly supported under NASA GI NNX08AJ20G and LWS TR\&T NNG0-5GN34G grants. P.G. and V.Y are partially supported by NSF (AGS-0745744) and NASA (NNY 08BA22G). Hinode is a Japanese mission developed and launched by ISAS/JAXA, with NAOJ as a domestic partner and NASA and STFC (UK) as international partners. It is operated by these agencies in co-operation with ESA and NSC (Norway).


\begin{thebibliography}{}
%
\bibitem[Bellot Rubio et al.(2004)]{Bel04}
    Bellot~Rubio,~L.~R., Balthasar,~H., \& Collados,~M. 2004, \aap, 427, 319
\bibitem[Bonet et al.(2005)]{Bon05}
    Bonet,~J.~A., M\'{a}rquez,~I., Muller,~R., Sobotka,~M., \& Roudier,~Th. 2005, \aap, 430, 1089
\bibitem[Bonet et al.(2004)]{Bon04}
    Bonet,~J.~A., M\'{a}rquez,~I., Muller,~R., Sobotka,~M., \& Tritschler,~A. 2004, \aap, 423, 737

\bibitem[Cabrera Solana et al.(2006)]{Cab06}
    Cabrera~Solana,~D., Bellot Rubio,~L.,~R., Beck,~C., \& Del~Toro~Iniesta,~J.~C. 2006, \apjl, 649, 41
\bibitem[Cabrera Solana et al.(2007)]{Cab07}
    Cabrera~Solana,~D., Bellot Rubio,~L.,~R., Beck,~C., \& Del~Toro~Iniesta,~J.~C. 2007, \aap, 475, 1067
\bibitem[Cabrera Solana et al.(2008)]{Cab08}
    Cabrera~Solana,~D., Bellot Rubio,~L.,~R., Borrero,~J.~M., \& del Toro Iniesta,J.C. 2008, \aap, 273, 283
\bibitem[Cao et al.(2010a)]{Cao10a}
    Cao,~W., Gorceix,~N., Coulter,~R., et al. 2010a, Astron.~Nachr., 331, 636
\bibitem[Cao et al.(2010b)]{Cao10b}
    Cao,~W., Gorceix,~N., Coulter,~R., et al. 2010b, SPIE., 7735, 77355V

\bibitem[Hagenaar \& Shine(2005)]{Hag05}
    Hagenaar,~H., \& Shine,~R.~A. 2005, \apj, 635, 659
\bibitem[Harvey \& Harvey(1973)]{Har73}
    Harvey,~K., \& Harvey,~J. 1973, \solphys, 28, 61

\bibitem[Ichimoto et al.(2008)]{Ich08}
    Ichimoto,~K., Lites,~B., Elmore,~D., et al. 2008, \solphys, 249, 233
\bibitem[Ichimoto et al.(2007)]{Ich07}
    Ichimoto,~K., Shine,~R.~A., Lites,~B., et al. 2007, \pasj, 59, S593

\bibitem[Katsukawa et al.(2011)]{Kat11}
    Katsukawa,~Y., Shimojo,~M., \& Yokoyama,~T. 2011, personal conversation
\bibitem[Kitiashvili et al.(2010)]{Kit10}
    Kitiashvili,~I.~N., Bellot~Rubio,~L.~R., Kosovichev,~A.~G., et al. 2010, \apjl, 716, L181
\bibitem[Kosugi et al.(2007)]{Kos07}
    Kosugi,~T., Matsuzaki,~K., Sakao,~T., et al. 2007, \solphys, 243, 3

\bibitem[Lee(1992)]{Lee92}
    Lee,~J.~W. 1992, \solphys, 139, 267
\bibitem[Lim et al.(2011)]{Lim11}
    Lim,~E.~-K., Yurchyshyn,~V., Abramenko,~V., et al. 2011, \apj, 740, 82

\bibitem[Moon et al.(2003)]{Moon03}
    Moon,~Y.~-J., Wang,~H., Spirock,~T.~J., Goode,~P.~R., \& Park,~Y.~D. 2003, \solphys, 217, 79
\bibitem[Mart\'{\i}nez Pillet(2000)]{Mar00}
    Mart\'{\i}nez Pillet,~V. 2000, \aap, 361, 734
\bibitem[Mart\'{\i}nez Pillet(2002)]{Mar02}
    Mart\'{\i}nez Pillet,~V. 2002, Astron.~Nachr., 323, 342

\bibitem[Ravindra(2006)]{Rav06}
    Ravindra,~B. 2006, \solphys, 237, 297
\bibitem[Rees \& Semel(1979)]{Ree79}
    Rees,~D.~E., \& Semel,~M.~D. 1979, \aap, 74, 1
\bibitem[Rempel(2011)]{Rem11}
    Rempel,~M. 2011, \apj, 729, 5
\bibitem[Ryutova et al.(1997)]{Ryu97}
    Ryutova,~M., Shine,~R., Title,~A., \& Sakai,~J.~I. 1997, \apj, 492, 402

\bibitem[Sainz Dalda \& Bellot Rubio(2008)]{Sai08}
    Sainz~Dalda,~A., \& Bellot Rubio,~R. 2008, \aap, 481, L21
\bibitem[Sainz Dalda \& Mart\'{\i}nez Pillet(2005)]{Sai05}
    Sainz~Dalda,~A., \& Mart\'{\i}nez Pillet,~V. 2005, \apj, 632, 1176
\bibitem[Schlichenmaier(2002)]{Sch02}
    Schlichenmaier,~R. 2002, Astron. Nachr., 323, 303
\bibitem[Schlichenmaier \& Schmidt(2000)]{Sch00}
    Schlichenmaier,~R., \& Schmidt,~W. 2000, \aap, 358, 1122
\bibitem[Semel(1967)]{Sem67}
    Semel,~M.~D. 1967, Ann. d'Astrophys., 30, 513
\bibitem[Semel(1970)]{Sem70}
    Semel,~M.~D. 1970, \aap, 5, 330
\bibitem[Sheeley(1969)]{She69}
    Sheeley,~N.~R. 1969, \solphys, 9, 347
\bibitem[Shimizu et al.(2008)]{Shi08}
    Shimizu,~T., Lites,~B.~W., Katsukawa,~Y., et al. 2008, \apj, 680, 1467
\bibitem[Shine \& Title(2001)]{Shi01}
    Shine,~R.~A., \& Title,~A.~M. 2001, in Encyclopedia of Astronomy and Astrophysics, ed. P. Murdin (Bristol: IOP Publishing), 3209
\bibitem[Shine et al.(1994)]{Shi94}
    Shine,~R.~A., Title,~A.~M., Tarbel,~T.~D., Smith,~K., \& Frank,~Z.~A. 1994, \apj, 430, 413
\bibitem[Shine et al.(1987)]{Shi87}
    Shine,~R.~A., Title,~A.~M., Tarbell,~T.~D., \& Topka,~K.~P. 1987, Science, 238, 1264
\bibitem[Solanki \& Montavon(1993)]{Sol93}
    Solanki,~S.~K., \& Montavon,~C.~A.~P. 1993, \aap, 275, 283
\bibitem[Stanchfield et al.(1997)]{Sta97}
    Stanchfield,~D.~C.~H.,~II, Thomas,~J.~H., \& Lites,~B.~W. 1997, \apj, 477, 485
\bibitem[Suematsu et al.(2008)]{Sue08}
    Suematsu,~Y., Tsuneta,~S., Ichimoto,~K., et al. 2008, \solphys, 249, 197

\bibitem[Title et al.(1993a)]{Tit93a}
    Title,~A.~M., Frank,~Z.~A., Shine,~R.~A., et al. 1993, \apj, 403, 780
\bibitem[Title et al.(1993b)]{Tit93}
    Title,~A.~M., Frank,~Z.~A., Shine,~R.~A., et al. 1993, \apj, 407, 398
\bibitem[Tsuneta et al.(2008)]{Tsu08}
    Tsuneta,~S., Suematsu,~Y., Ichimoto,~K., et al. 2008, \solphys, 249, 167

\bibitem[Uitenbroek(2003)]{Uit03}
    Uitenbroek,~H. 2003, \apj, 592, 1225

\bibitem[Vrabec(1974)]{Vra74}
    Vrabec,~D. 1974, in IAU Symp. 56, Chromospheric Fine Structure, ed. R.G.Athay (Dordrecht: Reidel), 201

\bibitem[Wang et al.(1991)]{Wan91}
    Wang,~H., Zirin,~H., \& Ai,~G. 1991, \solphys, 131, 53
\bibitem[Westendorp Plaza et al.(2001)]{Wes01}
    Westendorp~Plaza,~C., del~Toro~Iniesta,~J.~C., Ruiz~Cobo,~B., \& Mart\'{\i}nez~Pillet,~V. 2001, \apj, 547, 1148
\bibitem[Westendorp Plaza et al.(1997)]{Wes97}
    Westendorp~Plaza,~C., del~Toro~Iniesta,~J.~C., Ruiz~Cobo,~B., et al. 1997, Nature, 389, 47
\bibitem[W\"{o}ger et al.(2008)]{Wog08}
    W\"{o}ger,~F., von der L\"{u}he,~O., \& Reardon,~K. 2008, \aap, 488, 375

\bibitem[Yurchyshyn et al.(2001)]{Yur01}
    Yurchyshyn,~V.~B., Wang,~H., \& Goode,~P.~R. 2001, \apj, 550, 470

\bibitem[Zhang et al.(2003)]{Zha03}
    Zhang,~J., Solanki,~S.~K., \& Wang,~J. 2003, \aap, 399, 755
\bibitem[Zhang et al.(2007)]{Zha07}
    Zhang,~J., Solanki,~S.~K., Woch,~J., \& Wang,~J. 2007, \aap, 471, 1035
%
\end{thebibliography}
\end{document}